\documentclass[aps,amsfonts,pra,twocolumn,superscriptaddress]{revtex4-1}
\usepackage{epsfig,amsmath,amssymb,bm,epsf,graphicx,psfrag}
\usepackage[all]{xy}
\usepackage{color}
\usepackage{subfigure}

\def\Longarrow{\protect\@lra}
\def\@lra{\relbar\joinrel\relbar\joinrel\relbar\joinrel%
          \relbar\joinrel\rightarrow}

\newcommand{\bc}{\begin{center}}
\newcommand{\ec}{\end{center}}
\newcommand{\be}{\begin{equation}}
\newcommand{\ee}{\end{equation}}
\newcommand{\bea}{\begin{eqnarray}}
\newcommand{\eea}{\end{eqnarray}}

\newcommand{\ncd}{\newcommand}
\ncd{\QCcns}{$QC_{\cal{C}}$}
\ncd{\QCc}{$QC_{\cal{C}}\;$}

\definecolor{libl}{cmyk}{0.2,0.1,0,0}

\begin{document}

\title{Universal measurement-based quantum computation in two-dimensional SPT phases}
\author{Tzu-Chieh Wei}
\affiliation{C. N. Yang Institute for Theoretical Physics and
Department of Physics and Astronomy, State University of New York at
Stony Brook, Stony Brook, NY 11794-3840, USA}
\author{Ching-Yu Huang}
\affiliation{C. N. Yang Institute for Theoretical Physics and
Department of Physics and Astronomy, State University of New York at
Stony Brook, Stony Brook, NY 11794-3840, USA}
\affiliation{Physics Division, National Center for Theoretical Science, Hsinchu 30013, Taiwan}
\begin{abstract}
Recent progress in characterization for gapped quantum phases has also triggered the search of universal resource for quantum computation in symmetric gapped phases. Prior works in one dimension suggest that it is a  feature more common than previously thought  that nontrivial 1D symmetry-protected topological (SPT) phases provide quantum computational power characterized by the algebraic structure defining these phases. Progress in two and higher dimensions so far has been limited to special fixed points in SPT phases. Here we provide two families of 2D $Z_2$ symmetric wave functions such that there exists a finite region of the parameter in the SPT phases that supports universal quantum computation. The quantum computational power loses its universality at the boundary between the SPT and symmetry-breaking phases.   
\end{abstract}
\date \today
 \maketitle

 \section{Introduction and motivation}
 Quantum mechanics  allows certain computational
tasks to be performed much more efficiently than using classical rules.
The most celebrated example is the factoring of a large
integer by Shor's quantum algorithm~\cite{Shor} that offers exponential
speedup over existing classical algorithms, among many quantum algorithms showing superiority over classical counterparts~\cite{Grover,ChildsWalk,Hallgren,Harrow,BaconvanDam,Somma,Hillery,Love}. Quantum computers
that implement generic quantum algorithms can take various forms, such as the standard circuit
model~\cite{NielsenChuang}, the topological architecture~\cite{TQC}, the adiabatic quantum computation~\cite{Farhi,Averin}, and the quantum
walk framework~\cite{Childs2009}, all of which proceed via the important feature of
quantum mechanics---the unitary evolution, before reading out the result by measurement.

In contrast, the paradigm of  the measurement-based quantum computation (MBQC)~\cite{BriegelNatPhys,RaussendorfWei,Kwek}, with the teleportation-based
schemes~\cite{GottesmanChuang} and the one-way quantum computer~\cite{RaussendorfBriegel,RaussendorfBrowneBriegel,AliferisLeung,BriegelNatPhys,RaussendorfWei}  as the prominent examples,
offers an alternative framework, in which local measurement
alone achieves the same power of computation as other models, provided that
a prior sufficiently entangled state is given. 
One of the
challenges in MBQC is to identify these entangled states,
namely, the universal resource states that enable the success
of driving universal quantum computation. It is known that  states with too little entanglement, naturally,
cannot provide sufficient quantum correlation to drive universal
quantum computation~\cite{VandenNest1,VandenNest2}. However, if a state
possesses too much entanglement, the measurement outcome
is so random that cannot provide any advantage over classical random guessing~\cite{GrossFlammia,Bremner}. Thus, {it is the structure of the entanglement rather than its amount that is important for computation\/}.

\begin{figure*}[t]
\center{\epsfig{figure=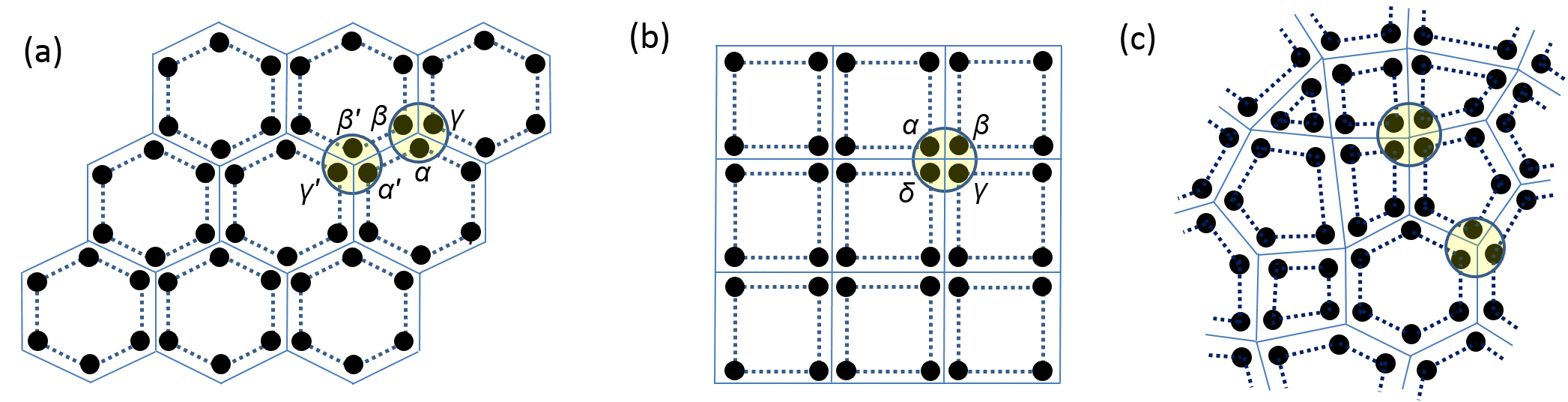,angle=0,width=0.92\textwidth}}
\caption{ Lattice and qubits: the each physical site (illustrated by a shaded circle) contains multiple qubits: three on (a) the honeycomb lattce, four on (b) the square lattice, and either three or four on (c) a random planar graph of mixed degrees 3 and 4. Each dashed loop indicates the constraint on the state of qubits connected by the loop; they need to the same either all $|0\rangle$ or all $|1\rangle$, i.e. the loop is a GHZ-like constraint: $|00\dots0\rangle+|11\dots1\rangle$.  The local weight on each site is represented by $A(\alpha,\beta,\gamma)$  on one sublattice or $A(\alpha',\beta',\gamma')$ the other sublattice of honeycomb [Eq.~(\ref{eqn:stateHoneycomb})] and by  $A[\alpha,\beta,\gamma,\delta]$ on the square lattice [Eq.~(\ref{eqn:stateSquare})]. These are straightforwardly extended to (c).}
  \label{fig:HoneycombSquare}
\end{figure*}

An intriguing connection of MBQC to condensed-matter physics emerges in the recent discovery of the symmetry-protected topological  (SPT) phases~\cite{SPTRG,1DSPTcomplete,PollmannSPT,SPTtensor,ChenScience,cohomology}, where states possess certain types of short-range entanglement. This connection was first revealed in the 1D SPT phase~\cite{ElseSchwarzBartlettDoherty}, where the SPT order can be utilized for the
protection of certain quantum gates in MBQC, even though
1D quantum states do not naturally accommodate universal quantum
computation. Examples include the 1D cluster state and the 1D spin-1 AKLT state, which can be used for arbitrary single-qubit gates~\cite{Gross,Brennen}. Such utility in quantum gates has been further  established~\cite{MillerMiyake,Abhi,David1,David2}. 

However, the development in two dimensions and higher is far limited, with only a handful of fixed-point wave functions providing universal resources~\cite{Hendrik,MillerMiyake16,MillerMiyake16b}, and the usefulness in some cases may depend on the underlying lattices. An important question remains open is whether the quantum computational universality can be extended beyond the fixed points in  two or high-dimensional SPT phases.
Here we consider two 2D families of symmetric wave functions and show that there exists a finite region in the parameter space that supports universal quantum computation. Such existence  does not depend on lattices. The quantum computational power diminishes at the boundary of the SPT and symmetry-breaking phases.   This shows positively that quantum computational universality is a feature beyond just the fixed point and is strongly related to the SPT order. (We remark that the universality used in this paper refers to quantum computation rather than the universality class in phase transitions.)

 \section{Families of $Z_2$ symmetric wave functions: SPT and symmetry-breaking phases }
 Here we present two families of $Z_2$ symmetric wave functions on the honeycomb and square lattices, respectively. 
 Each physical site contains multiple qubits and each qubit forms a GHZ-like loop,  $|00\dots0\rangle+|11\dots1\rangle$,  with other qubits on a face of each lattice. These GHZ loops form the ground state of the so-called CZX model~\cite{CZX}, but we generalize the construction beyond this fixed-point model. Our wave functions can be regarded as deformation from these exact GHZ loops and the weight of deformation (denoted by $A$) depends on the spin configurations on each site. Referring to Fig.~\ref{fig:HoneycombSquare} for the arguments of the tensor $A$, we have on the honeycomb lattice (previously studied by us~\cite{HuangWei}):
 \begin{align}
& A(0,0,0) = A(1,1,1) = 1 \notag \\
& A(0,0,1) = A(0,1,0) = A(1,0,0) = |g| \notag \\
& A(1,1,0) = A(1,0,1) = A(0,1,1) = g,
\label{eqn:stateHoneycomb}
\end{align}
and on the square lattice:
\begin{align}
\label{eqn:stateSquare}
& A[0,0,0,0] = A[1,1,1,1] = 1 \\
& A[0,0,1,1] = A[1,0,0,1]=g \notag \\ 
& A[1,1,0,0] = A[0,1,1,0]= A[0,1,0,1] = A[1,0,1,0] = |g|  \notag \\
& A[0,0,1,0] = A[1,1,0,1] = A[1,0,0,0] = A[0,1,1,1] = |g|  \notag \\
& A[0,1,0,0] = A[0,0,0,1]  = A[1,0,1,1] = A[1,1,1,0] = |g|\notag.
\end{align}
As one can see that under the local $Z_2$ action $\hat{u}=\sigma_x\otimes \dots \otimes\sigma_x$, these weights display a symmetry, up to a possible sign (if $g<0$), showing that the wave functions are symmetric under the action of the simple $Z_2$ group $G=\{\openone, \hat{u}\}$, whose action  can also be understood in terms of the so-called matrix-product operators (see Appendix~\ref{app:MPO}). Short-range gapped parent Hamiltonians can be constructed such that these wave functions are the ground states~\cite{HuangWei}.

\begin{figure}[t]
\includegraphics[width=0.47\textwidth]{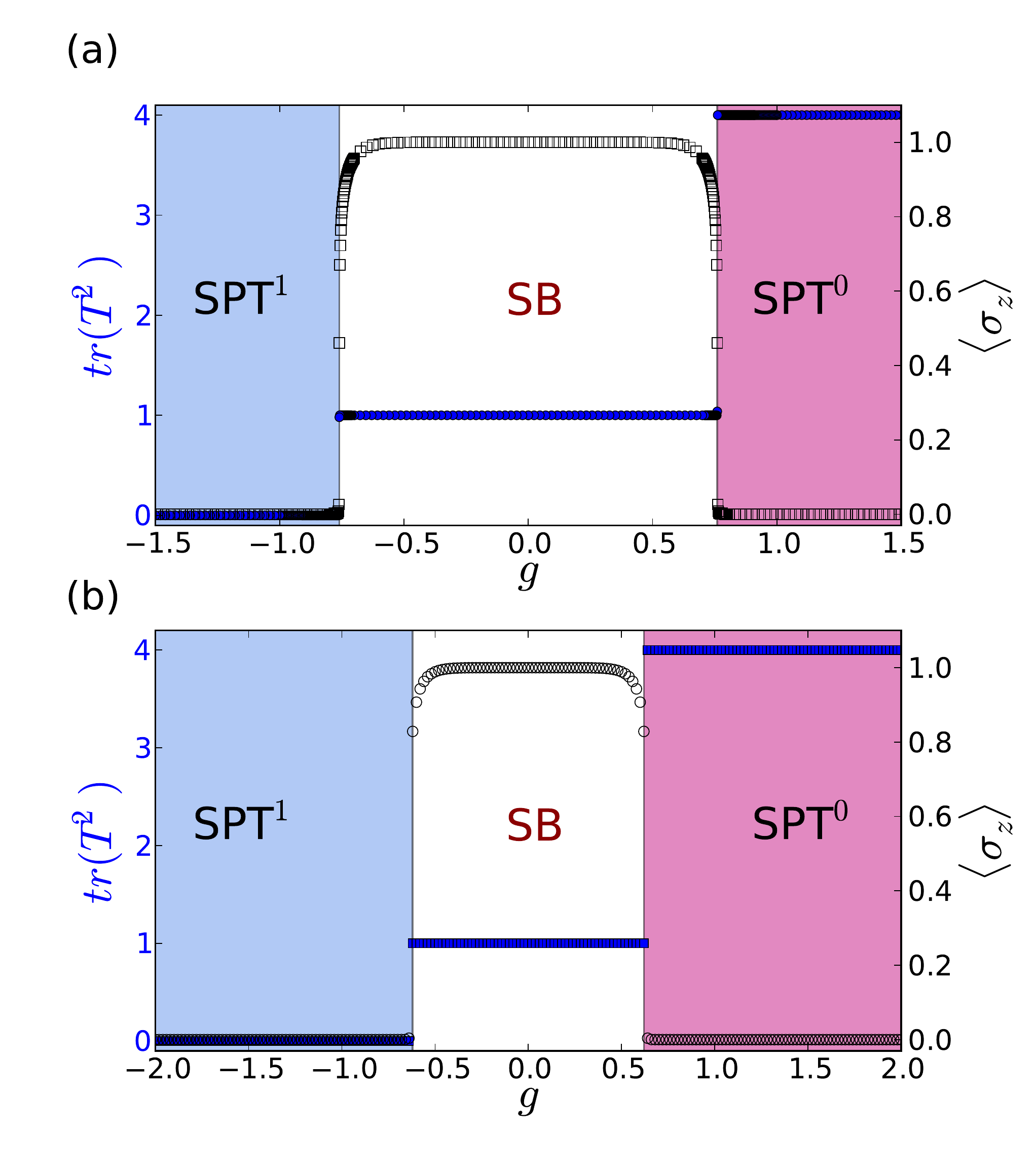}
\caption{Probing the phase diagram vs. $g$ using the simulated modular T matrix and the local order parameter. ${\rm Tr}(T^2)$ is represented by black empty squares and the order parameter $\langle \sigma_z\rangle$ is represented by blue empty circles. (a) Top panel: honeycomb lattice (see also Ref.~\cite{HuangWei}); (b) bottom panel: square lattice. The transitions are estimated to be at $|g_c|=0.759(1)$ on the honeycomb lattice and at $|g_c|=0.633(1)$ for the square lattice.}
\label{fig:SandT_matrix}
\end{figure} 
Inspired by the works of Levin and Gu and of Hung and Wen~\cite{LevinGu2012,HungWen}, we developed a tensor-network method (referred to as tnST) to probe the SPT phases by using the simulated modular matrices $S$ and 
$T$~\cite{HuangWei}, which originate from the duality between SPT and intrinsic topological order.  Such a tnST method can distinguish nontrivial SPT order from trivial SPT order and spontaneous symmetry breaking by examining modular matrices. Essentially for the $Z_2$ symmetry, only the trace of modular $T^2$ is needed for the characterization of the SPT order~\cite{HungWen}, and we show the results in Fig.~\ref{fig:SandT_matrix} that depict several distinct phases. These two families of $Z_2$ symmetric wave functions display a common feature in their phase diagram: for the parameter $g$ sufficiently negative, it is a nontrivial SPT phase (labeled as SPT${}^1$), and as $g$ increases, there is a transition to a $Z_2$ spontaneous symmetry-breaking phase (labeled as SB), followed by another transition to a trivial SPT phase (labeled as SPT${}^0$).   The detailed forms of modular $S$ and $T$ matrices in different regions are listed in Appendix~\ref{app:ST}. In the symmetry-breaking phase, the local order parameter is $\langle \sigma_z\rangle$ and it vanishes at the same boundaries obtained from the modular matrices.

\section{Universal quantum computation in SPT phases}
In a prior work, one of us has shown that universal quantum computation can be done using fixed-point nontrivial SPT states of Chen,  Gu, Liu and Wen~\cite{cohomology} (for any group where the 3rd group cohomology is nontrivial) in the framework of MBQC on any 2D lattices~\cite{Hendrik}. Miller and Miyake also provide  examples of $Z_2^3$ SPT-symmetric fixed-point wave function on the union-jack lattice~\cite{MillerMiyake16,MillerMiyake16b}. But an important question of whether there exists an extended region (if not the whole phase) in an SPT phase that supports universal MBQC. As we shall see the two families of symmetric wave functions provide an affirmative answer;  indeed, in a finite region of the parameter $g$, universal quantum computation is supported. To do this, we will construct local generalized measurement (i.e. POVM) so that we can convert the deformed wave function back to a fixed point, albeit the effective lattice structure (for which the GHZ loops are concerned) is modified. By indentifying the range of $g$ such that the number of GHZ loops is macroscopic and no macroscopic size of GHZ loops, we can narrow down the quantum computational universality region. Moreover, we find that as one approaches the symmetry-breaking phase from the SPT side (either nontrivial or trivial), the universal quantum computational capability diminishes at the transition.

For simplicity we will focus on the honeycomb case, but our results also hold for the square case (relegated to the Appendix~\ref{app:Square}), except that the range of universality is different. Since the sign of $g$ is local to each site of three qubits, for the purpose of quantum computation we can transform it away by a local unitary (or equivalently a local basis change):
\begin{eqnarray}
U(g)&=&|000\rangle\langle 000|+|111\rangle\langle 111|\\ 
& &+|001\rangle\langle001|+|010\rangle\langle010|+|100\rangle\langle100|\nonumber\\
&& +{\rm sgn}(g)(|110\rangle\langle110|+|101\rangle\langle101|+|011\rangle\langle011|), \nonumber
\end{eqnarray}
where ${\rm sgn}(g)$ is the sign function, which we take to be 1 if $g\ge 0$, and -1 if $g<0$. Therefore, it sufficies to take $g\ge 0$ and divide our consideration to two cases: (i) $g\le 1$ and (ii) $g>1$ and our results will hold for $g<0$, too.
\begin{figure}
\center{\epsfig{figure=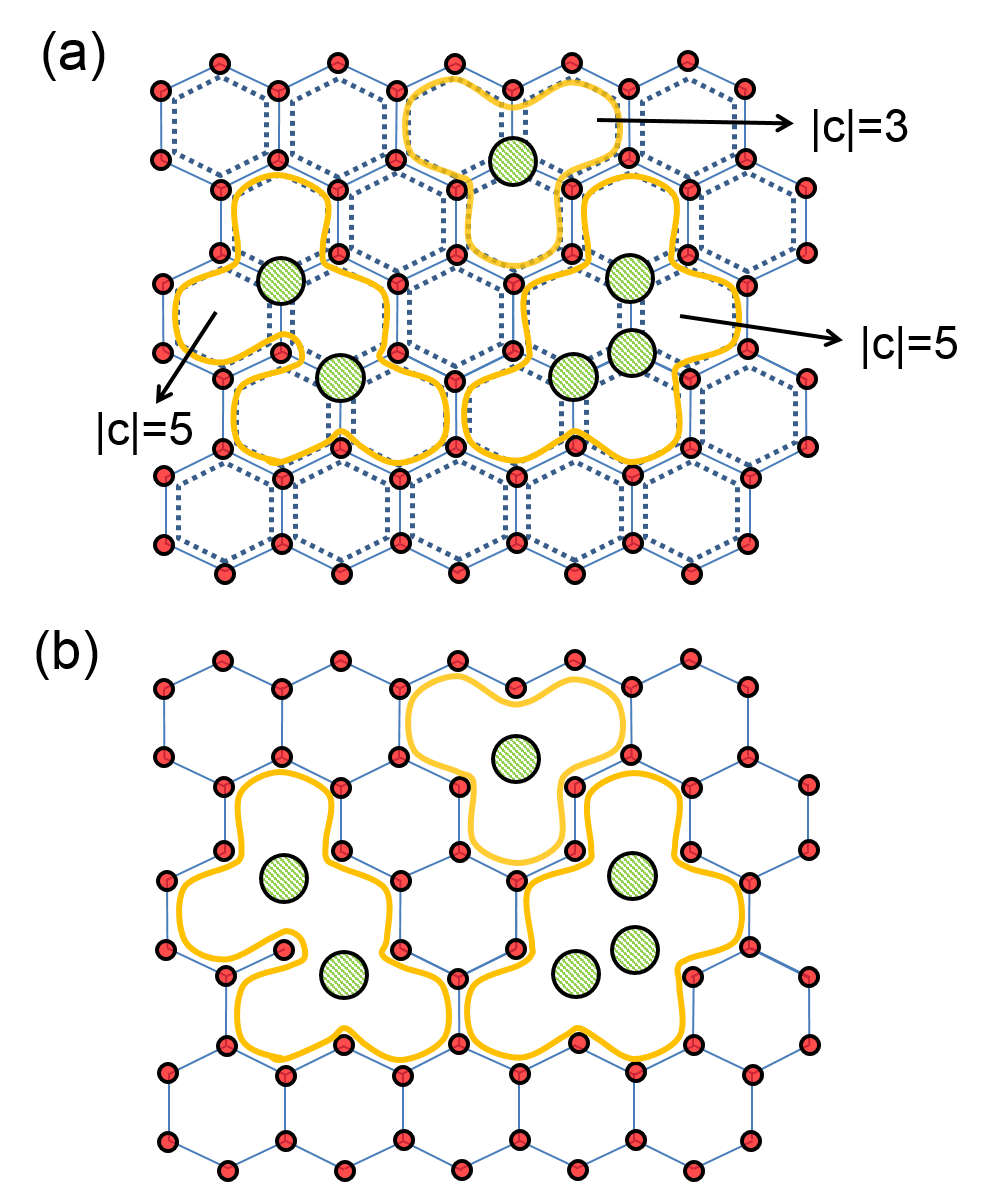,angle=0,width=0.48\textwidth}}
\caption{ Illustration of  POVM outomes and GHZ loops. Solid green (larger, meshed) dots represent outcomes of $E_2$, while smaller red dots represent $E_1$ outcomes.  Individual qubits are not shown. (a) A cluster is defined as a loop composed of one or multiple hexagons, with those having multiple hexagons indicated by a curved loop enclosing them. The sizes in the example clusters are indicated by $|c|$. (b) A simplified rendition of (a). }
  \label{fig:loops}
\end{figure}

\subsection{Case $g\le 1$}
We construct a  local  generalized measurement (i.e. POVM) that contains two elements $\{ E_1=F_1^\dagger F_1, E_2=F_2^\dagger F_2\}$, where
\begin{eqnarray}
F_1&=&g(|000\rangle\langle 000|+|111\rangle\langle 111|)\nonumber\\ 
& &+|001\rangle\langle001|+|010\rangle\langle010|+|100\rangle\langle100|\nonumber\\
&& +|110\rangle\langle110|+|101\rangle\langle101|+|011\rangle\langle011|, \\
F_2&=&\sqrt{1-g^2}(|000\rangle\langle 000|+|111\rangle\langle 111|).
\end{eqnarray}
One can verify that $E_1+E_2=\openone$ and thus the POVM is trace-preserving. The meaning of this POVM is as follows. Depending on the outcome $i =1{\,\rm or\,\,}2$, an initial state $|\psi\rangle$ is mapped to $F_i|\psi\rangle$ after the measurement. When the outcome is $E_1$, the local tensor is reduced back to the fixed-point form, except for the $g$ factor: $A(\alpha,\beta,\gamma)=g$.
  When the outcome is $E_2$, the local tensor becomes
  \begin{equation}
A(0,0,0) = A(1,1,1) = \sqrt{1-g^2}, \ A({\rm rest})=0,
\end{equation}
where `rest' indicates all other configurations than $(0,0,0)$ and $(1,1,1)$. Thus, the $E_2$ outcome results in the merging of the three GHZ loops surrounding the vertex into one giant loop; see Fig.~\ref{fig:loops}. All qubits on and enclosed by the giant loop form a GHZ state.

 \begin{figure}
\subfigure[]{\includegraphics[width=0.4\textwidth]{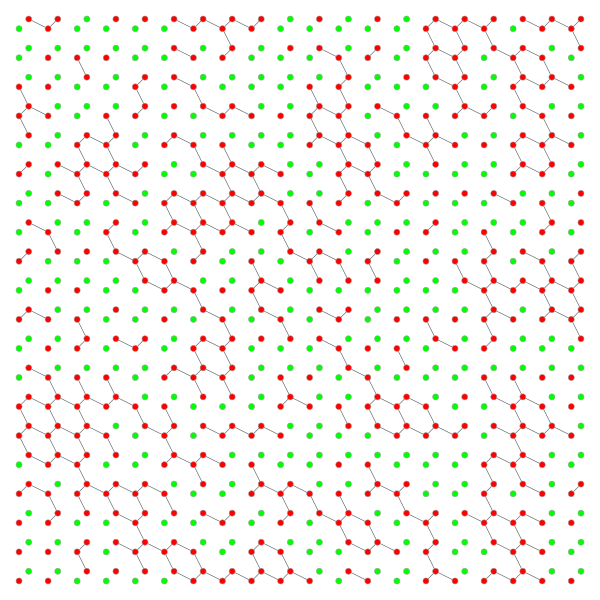}}
\subfigure[]{\includegraphics[width=0.4\textwidth]{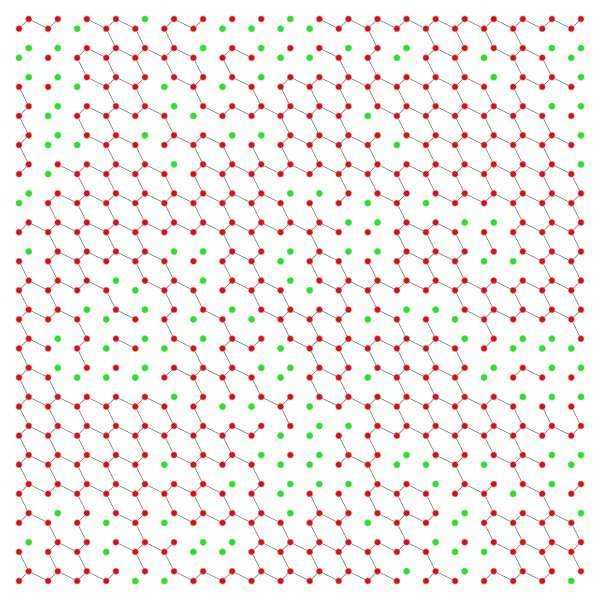}}
\caption{ Illustration of  POVM outomes of a honeycomb lattice of 800 hexagons; (a) top panel: $g=0.74$ (a macroscopic loops) and (b) bottom panel: $g=0.78$ (no macroscopic loops). Loops are suppressed and $E_1$ outcomes are indicated by red dots (darker shades) and $E_2$ outcomes by green dots (lighter shades). But existence of a macrosopic loop can be seen from the existence of a pathway (like in a maze) from the left side to the right one or from the top to the bottom.  }
  \label{fig:graphloops}
\end{figure}
 
\begin{figure}[t]
\center{\epsfig{figure=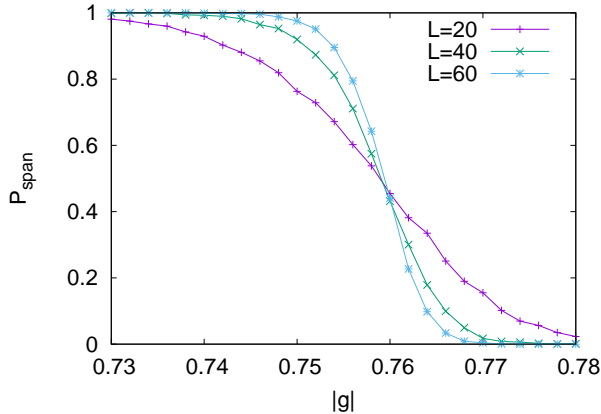,angle=0,width=.5\textwidth}}
\vspace{-0.8cm}
\caption{Probability $P_{\rm span}$ of a spanning loop (i.e. a macroscopic GHZ loop) vs. parameter $g$ for the case $g<1$ for various linear sizes $L$ on the honeycomb lattice. The total number of hexagons is $2L^2$.  Each data point is averaged over 2000-4000 samples and the statistical error is less than 1\%. A percolation transition is estimated at $g_{c1}\approx 0.760(2)$ in the thermodynamic limit. In the region $g_{c1}< g\le 1$, there is no macroscopic GHZ loop ($P_{\rm span}\rightarrow 0$ as $L\rightarrow \infty$) and hence a sufficiently large valence-bond state with the number of qubits being proportional to the original system size can be obtained. This means that the system provides a universal resource for MBQC.}
\label{fig:perco}
\end{figure}

 We label the POVM outcome by $\{\alpha_{v}\}$, where $v$ denote a site and $\alpha_v$ can be either $1$ (desired) or $2$ (undesired) measurement outcome. 
 We label the total number of outcome $E_1$ by $n_1$ and that of $E_2$ by $n_2$, as well as the size of cluster $c$  of loops by $|c|$ that counts the number of hexagons in a GHZ loop (see Fig.~\ref{fig:loops}). An initial state $|\psi\rangle$ is transformed by the measurement to $\mathop{\otimes}_v F_{\alpha_v} |\psi\rangle$, and the probability that  $\{\alpha_v\}$ occurs, by definition, is 
  \begin{equation}
P_{\{\alpha_v\}}(g)=\frac{\langle\psi(g)|\mathop{\otimes}_v E_{\alpha_v} |\psi(g)\rangle}{\langle \psi(g)|\psi(g)\rangle}=\frac{\big\Vert\mathop{\otimes}_v F_{\alpha_v} |\psi\rangle\big\Vert^2}{\langle \psi(g)|\psi(g)\rangle}.
 \end{equation}
 This is a many-body correlation function, but we show in Appendix~\ref{app:Honeycomb1} that the result is 
 \begin{equation}
 \label{eqn:P}
 P_{\{\alpha_v\}}(g)=p_0\, g^{2n_1} (1-g^2)^{n_2}\, 2^{\sum_c (1-|c|)},
 \end{equation}
 where $p_0$ is an overall constant independent of $\{\alpha_v\}$.  Such an expression can be interpreted as the Boltzmann weight of a statistical model of interacting particles of two types; see Appendix~\ref{app:Honeycomb1}.
 With Eq.~(\ref{eqn:P}), we can employ the standard Monte Carlo method to efficiently sample important configurations $\{\alpha_v\}$ for statistical analysis of, e.g., graph properties (see Appendix~\ref{app:MonteCarlo} for details). From each outcome configuration, as illustrated in Fig.~\ref{fig:graphloops}, we examine whether an undesired macroscopic loop may exist, using a percolation approach for probing the probability of such a spanning loop. As shown in Fig.~\ref{fig:perco}, for $g>g_{c1}\approx 0.760(2)$, no such spanning loop exists in the thermodynamic limit, and thus there will be macroscopic number of small loops that can be further converted by additional local measurements to a valence-bond state which is universal~\cite{Hendrik,Verstraete}. We find that the location of $g_{c1}\approx 0.760(2)$ to be very close, if not identical, to the transition from the SPT phase to the $Z_2$ symmetry-breaking phase.  This is due to the emergence of long-range order as $g$ approaches $g_{c1}$.

 \subsection{Case $g> 1$}
 In this region, we need to employ a different 
 trace-perserving POVM that contains two elements $\{ E_3=F_3^\dagger F_3, E_4=F_4^\dagger F_4\}$, where
\begin{eqnarray}
\!\!\!\!\!\!\!F_3&=&\frac{1}{g}(|001\rangle\langle001|+|010\rangle\langle010|+|100\rangle\langle100|\nonumber\\
 \!\!\!\!\!\!\!& & +|110\rangle\langle110|+|101\rangle\langle101|+|011\rangle\langle011|)\\
& &+(|000\rangle\langle 000|+|111\rangle\langle 111|)\\
\!\!\!\!\!\!\!F_4&=&{\frac{\sqrt{g^2-1}}{g}}(|001\rangle\langle001|+|010\rangle\langle010|+|100\rangle\langle100|\nonumber\\
\!\!\!\!\!\!\!& & +|110\rangle\langle110|+|101\rangle\langle101|+|011\rangle\langle011|).
\end{eqnarray}
 When the outcome is $E_3$, the local tensor is reduced back to the fixed-point form, except for the $1/g$ factor. When the outcome is $E_4$, the local tensor becomes
 \begin{align}
 A(0,0,0) = A(1,1,1) =0, \ 
 A({\rm rest}) = \sqrt{g^2-1},
\end{align}
and this means that the three GHZ loops surrounding the vertex have been merged into one giant loop of a generalized GHZ state that can have more than two components. For example, one single site with $E_4$ outcomes merged the three neighboring hexagon loops of GHZ states into one GHZ state of six components 
\begin{align}
& |0...0\rangle|0...0\rangle|1...1\rangle+ |1...1\rangle|1...1\rangle|0...0\rangle \nonumber\\
&+ |0...0\rangle|1...1\rangle|0...0\rangle+|1...1\rangle|0...0\rangle|1...1\rangle \nonumber\\
&+ |1...1\rangle|0...0\rangle|0...0\rangle+|0...0\rangle|1...1\rangle|1...1\rangle.
\end{align}
 To recover the case of two-component GHZ state, one can follow up with a local measurement that consists of four projections  at the vertex (after the $E_4$ outcome): $P_0=|000\rangle\langle000|+|111\rangle\langle111|$, $P_1=|001\rangle\langle001|+|110\rangle\langle110|$, $P_2=|010\rangle\langle010|+|101\rangle\langle101|$, $P_3=|100\rangle\langle100|+|011\rangle\langle011|$. There will not be a result associated with $P_0$, and each outcome associated with either $P_1$, $P_2$ or $P_3$, projects the local Hibert space to a two-level subspace that can be locally converted to the $P_0$ subspace. The reduction to a universal resource state then goes as before.

 We use a similar labeling of the POVM outcome by $\{\alpha_{v}\}$, where $\alpha_v$ can be either $3$ (desired) or $4$ (undesired), and the total number of outcome $E_3$ by $n_3$ and that of $E_4$ by $n_4$. It is easy to see that (see also Appendix~\ref{app:Honeycomb2})
 \begin{equation}
 \label{eqn:P'}
 P_{\{\alpha_v\}}=p_0'\, g^{-2n_3} \left(\frac{g^2-1}{g^2}\right)^{-n_4} \, \prod_{c} q_c\, 2^{-|c|} ,
 \end{equation}
 where $p_0'$ is an overall constant that is independent of $\{\alpha_v\}$, and $q_c$ counts the number of components in the merged generalized GHZ state for the cluster $c$. For the previous $g<1$ case, the $q_c=2$, but due to the more complicated structure in the $E_4$, here, $q_c$ can be larger than 2.  We do not know of any exact expression for $q_c$ but we can bound the $q_c$ by $6\le q_c\le 6^{|c|/3}$, and these can be used to find the bounds of the corresponding transition point $g_{c2}$, for which we obtain $1.205(5)\lesssim g_{c2} \lesssim 1.390(2)$; see Appendix~\ref{app:MonteCarlo}.
 
  However, there is no transition of phases nearby. There are two possibilities: one being that our POVM for $g>1$ is not optimal but the quantum computational power extends further beyond, or it is possible that the quantum computational power does not extend all the way to the whole SPT phase. However, at the present we do not have a means to resolve this. 

\subsection{Square lattice}
For the $Z_2$ symmetric wave functions on the square lattice, we repeat a similar procedure and obtain that there also exists a finite region in the SPT phases: $g_{c1}< |g|<g_{c2}$ supporting universal quantum computation, and we estimate that 
 $g_{c1}\approx 0.635(3)$ and
 $1.31(1)\lesssim g_{c2}\lesssim 1.82(1)$; see also Appendix~\ref{app:Square}.
 
\subsection{Breaking translational invariance}
The construction above can be straightforwardly extended to other trivalent lattices, such as the square octagon, cross and star lattices and other four-valent lattices, such as the kagome lattice,
and the probablility distribution of the POVM outcomes will of similar forms. At $|g|=1$ the wave functions are the fixed-point ones and thus the quantum computational universality is guaranteed. Away from $|g|=1$, the probability of undesired POVM outcomes, such as $E_2$ or $E_4$, is small so there will be a finite region around $|g|=1$ such that the universality persists. The range of $g$ can be found using methods described here but depends on the lattice and is a generalized percolation problem. 
To break the translational invariance, we can consider random planar graphs with vertex degrees being three or four or even of mixed degree (see Fig.~\ref{fig:HoneycombSquare}c) and construct the family of wave functions in a similar way.   
As long as these graphs reside in the supercritical phase of percolation, the fixed point  $|g|=1$ possesses quantum computational universality. Away from $|g|=1$ as in the uniform case, there is a finite region around it such that the universality is perserved.  Thus our results of a finite region in the $Z_2$ SPT phase with quantum computational universality do not rely on translation invariance.  

 \section{Discussions}
 
We have constructed families of $Z_2$ symmetric wave functions and identified three distinct phases using modular matrices. For an extended region in the SPT phases, we have shown that universal quantum computation is possible via local measurements. Our construction can be straightforwardly extended to the non-translation invariant case. Previous quantum computational universality was only known for certain SPT fixed-point wave functions and may also depend on the lattice. Our results go beyond those and strengthen the potential connection of SPT order and quantum computation, which has been much explored in one dimension.

From the viewpoint of SPT order and higher dimensionality, the only exception so far  is the family of  AKLT states in two and higher dimensions~\cite{WeiAffleckRaussendorf11,Miyake11,WeiAffleckRaussendorf12,Wei13,WeiEtAl,WeiRaussendorf15} and many of them have been shown to provide  universal resource for quantum computation even beyond the AKLT points~\cite{DarmawanBrennenBartlett,Darmawan2,HuangWagnerWei}.  But their SPT order requires translation invariant symmetry to be respected. For our $Z_2$ symmetric wave functions, the existence of SPT order and the quantum computational universality do not require translation invariance to be respected.

It may seem puzzling why both the trivial and nontrivial SPT phases in our construction share similar quantum computational capability. This is because of our choice of symmetry group that protects the phase. To claim a state to be in an SPT phase, we need to specify the symmetry group. Even though our $g>0$ families of states belong to a trivial $Z_2$ SPT phase with respect to the simple $Z_2$ group used here, they actually display nontrivial SPT order with respect to another $Z_2$ group generated by the so-called CZX action~\cite{CZX}. But then for the full-range of the parameter $g$ the symmetry action depends on the sign of $g$, which is not a desirable feature. In our construction, the $Z_2$ action is the same across the whole parameter range, and is simply a spin flip.

Our results can be extended to three dimensions. However, the fourth cohomology group of $Z_2$ is trivial, so to have a nontrivial SPT phases, we need to have the symmetry group larger than $Z_2$, such as $Z_2\otimes Z_2$. Then the qubit in our $Z_2$ example needs to be replaced by a qu-quart, i.e. 4-level entity in our construction.

From a very different perspective, our results also imply that classical simulations of local measurements on the $Z_2$ SPT phases can be as hard as simulating a universal quantum computer, i.e. BQP-hard. But on the other hand simulating only the POVM outcomes on these  wave functions, such as those constructed and demonstrated here, is easy, i.e. in P.

\medskip \noindent {\bf Acknowledgment.}  This work was supported by the
National Science Foundation under Grants No. PHY 1314748 and No. PHY
1620252.

\bigskip
\appendix

   In this Appendix, we supply detailed discussions of points that were mentioned in the main text. Even though their omission does not affect the understanding of our main results, a more thorough exposition can still be useful for further development. 
 
\section{ $Z_2$ symmetry of the wave functions} 
\label{app:MPO}
\subsection{Honeycomb lattice}
\begin{figure*}[ht]
\center{\includegraphics[width=0.9\textwidth]{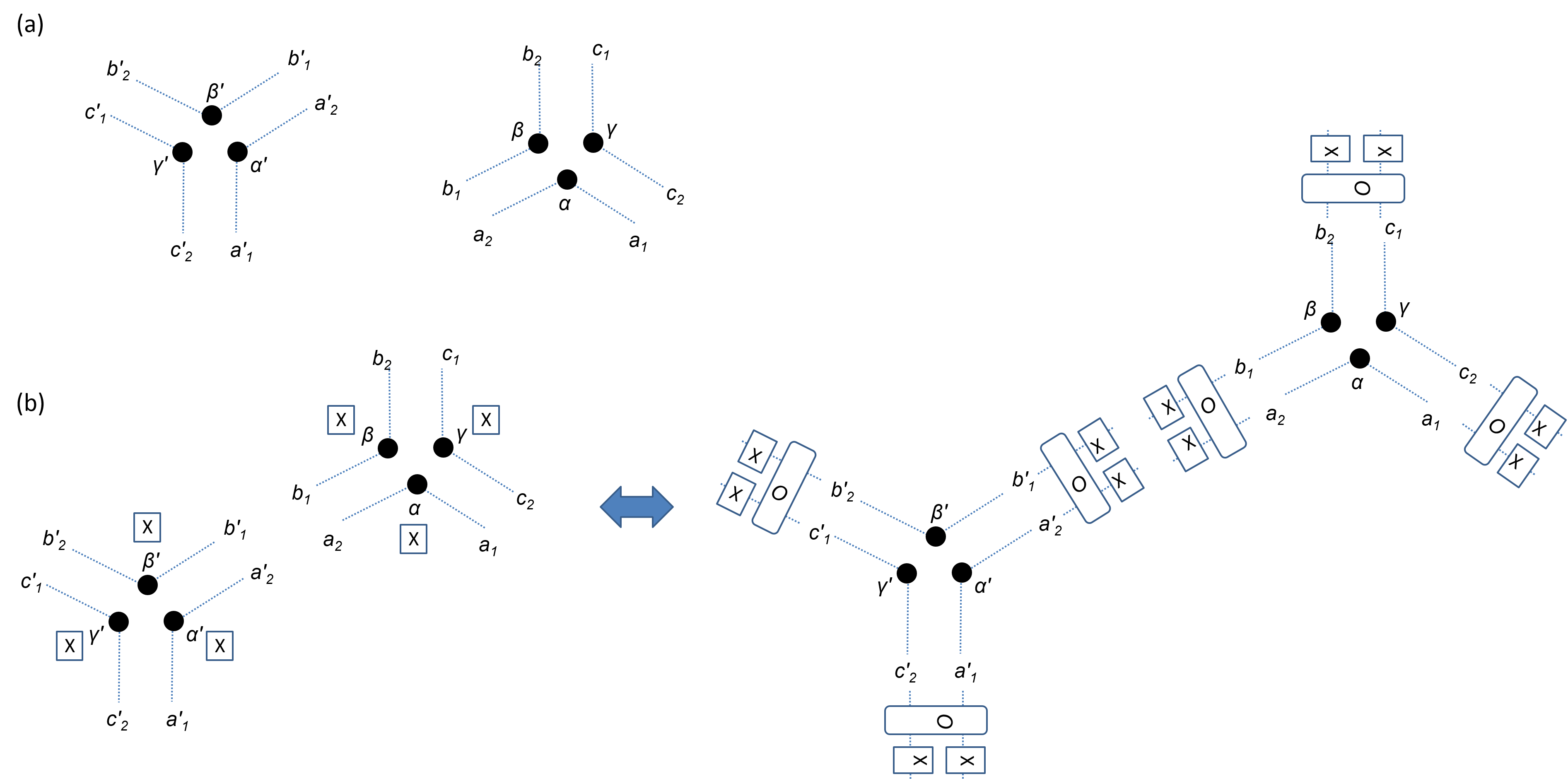}}
\caption{ (a) Local tensors for the wave functions. (b) The $Z_2$ symmetry action on the physical spins is equivalent to the local MPO action in the inner indices. In the drawing, operator $\hat{O}$ is applied first, followed by $X\otimes X$. But in the equation for the local MPO, one usually writes $\hat{O}\, X\otimes X$, reversing the order. As one can see that the neighboring local MPOs cancel and for a closed surface the wave functions are indeed $Z_2$ symmetric. }
  \label{fig:MPOHoneycomb}
\end{figure*}

For convenience we reproduce the wave functions on the honeycomb lattice as follows,
\begin{align}
& A(0,0,0) = A(1,1,1) = 1 \notag \\
& A(0,0,1) = A(0,1,0) = A(1,0,0) = |g| \notag \\
& A(1,1,0) = A(1,0,1) = A(0,1,1) = g.
\label{app:stateHoneycomb}
\end{align}
The wave functions can be represented by local tensors, as shown in Fig.~\ref{fig:MPOHoneycomb}(a), where the indices for the physical spin degree of freedom, e.g., $\alpha$, is indentical to the two inner indices, $a_1$ and $a_2$, which originates from the constraint of an underlying GHZ entanglement.
It is easy to verify that these wave functions are invariant under the $\mathbb{Z}_2$ action generated by
 the operator ${X} = \sigma_x= |0\rangle\langle 1|+|1\rangle\langle 0|$ (indicated on a boxed $X$ next to each physical index) on all partons.
However, the transformed tensors differ from the original ones by a local unitary transformation or local matrix-product operator (MPO) on the inner indices which is given by $ X \otimes X \hat{O}$, as shown in Fig.~\ref{fig:MPOHoneycomb}(b), where 
\begin{align}
\hat{O} =  | 00\rangle \langle 00 | +  | 01\rangle \langle 01 | +{\rm sgn}(g)| 10\rangle \langle 10 | + | 11\rangle \langle 11 |,
\end{align}
and ${\rm sgn}(g)$ is the sign function, which we take to be 1 if $g\ge 0$, and -1 if $g<0$.
This also gives a way of verifying that the wave function is $\mathbb{Z}_2$ symmetric, as the MPOs from neighboring sites cancel one another.

For such a fixed-point SPT wave function and others for nontrivial SPT phases, it was shown by us~\cite{Hendrik} that they provide universal resource for quantum computation in the measurement-based model, where only local operations or measurements are needed to achieve universal quantum computation.

\subsection{Square lattice}
\begin{figure}[h]
\center{\includegraphics[width=0.48\textwidth]{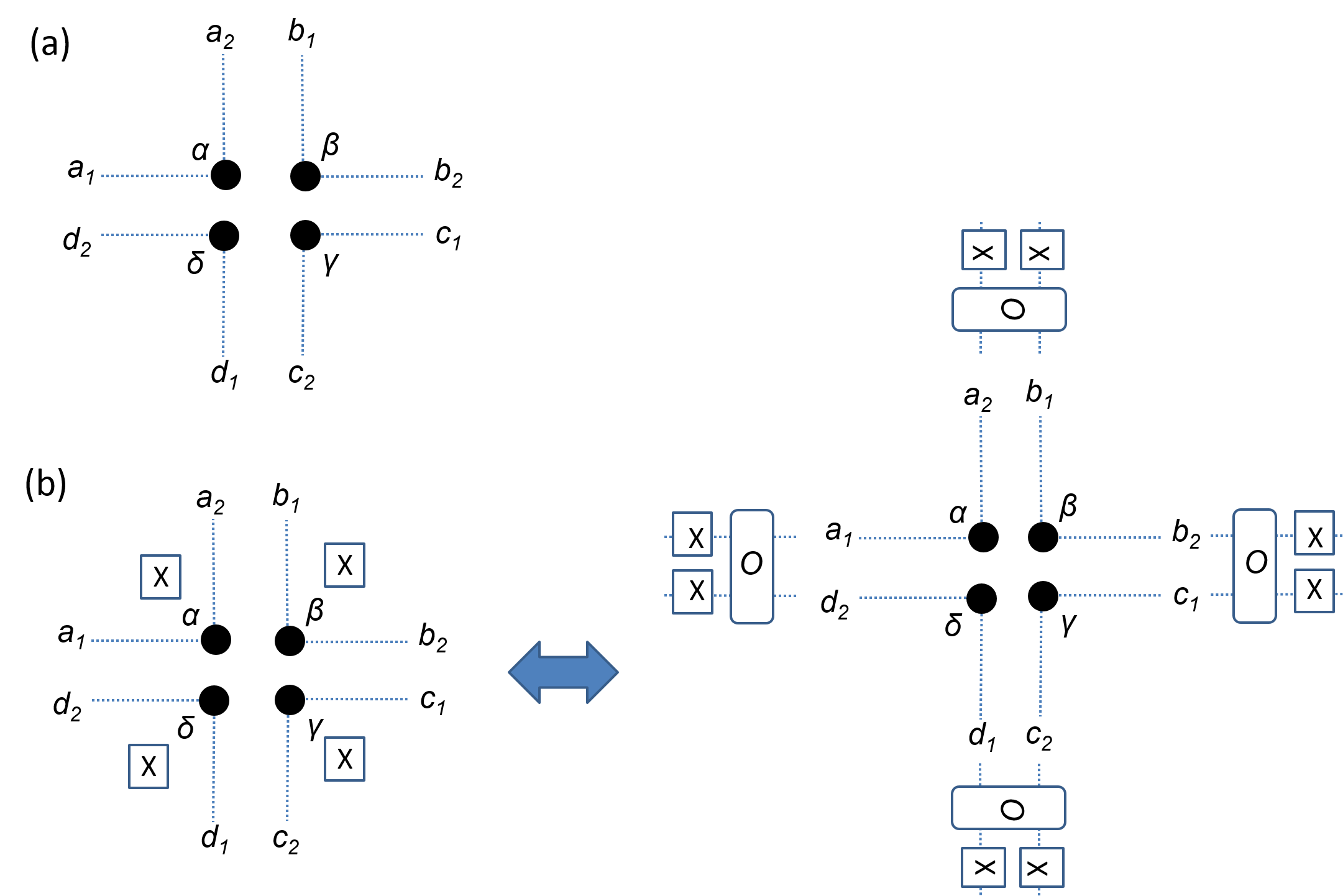}}
\caption{ (a) Local tensors for the wave functions. (b) The $Z_2$ symmetry action on the physical spins is equivalent to the local MPO action in the inner indices. In the drawing, operator $\hat{O}$ is applied first, followed by $X\otimes X$. But in the equation for the local MPO, one usually writes $\hat{O}\, X\otimes X$, reversing the order. As one can see that the neighboring local MPOs cancel and for a closed surface the wave functions are indeed $Z_2$ symmetric. }
  \label{fig:MPOSquare}
\end{figure}

For convenience we reproduce the wave functions on the square lattice as follows,
\begin{align}
\label{app:stateSquare}
& A[0,0,0,0] = A[1,1,1,1] = 1 \\
& A[0,0,1,1] = A[1,0,0,1]=g \notag \\ 
& A[1,1,0,0] = A[0,1,1,0]= A[0,1,0,1] = A[1,0,1,0] = |g|  \notag \\
& A[0,0,1,0] = A[1,1,0,1] = A[1,0,0,0] = A[0,1,1,1] = |g|  \notag \\
& A[0,1,0,0] = A[0,0,0,1]  = A[1,0,1,1] = A[1,1,1,0] = |g|\notag.
\end{align}
Similar to the honeycomb case, the wave functions on the square lattice can be represented by local tensors, as shown in Fig.~\ref{fig:MPOSquare}(a).
It is easy to verify that these wave functions are invariant under the $\mathbb{Z}_2$ action generated by
 the operator ${X} = \sigma_x= |0\rangle\langle 1|+|1\rangle\langle 0|$ (indicated on a boxed $X$ next to each physical index) on all partons.
However, the transformed tensors differ from the original ones by a local unitary transformation or local matrix-product operator (MPO) on the inner indices which is given by $ X \otimes X \hat{O}$, as shown in Fig.~\ref{fig:MPOSquare}(b), where  $\hat{O}$ differs slightly from that in the honeycomb case,
\begin{align}
\hat{O} =  | 00\rangle \langle 00 | +  | 01\rangle \langle 01 | +| 10\rangle \langle 10 | +{\rm sgn}(g) | 11\rangle \langle 11 |.
\end{align}
This also gives a way of verifying that the wave function is $\mathbb{Z}_2$ symmetric, as the MPOs from neighboring sites cancel one another.

\section{Probability of a POVM outcome: the honeycomb lattice}
Here we derive the probability for the POVM outcome, labelled by $\{\alpha_v\}$, on all sites of the honeycomb lattice.
\subsection{Case $|g|\le 1$}
\label{app:Honeycomb1}
Since the phase is local to each site of three qubits, for the purpose of quantum computation we can transform it away by local unitary. We can apply a local unitary transformation on each site: 
\begin{eqnarray}
U(g)&=&|000\rangle\langle 000|+|111\rangle\langle 111|\\ 
& &+|001\rangle\langle001|+|010\rangle\langle010|+|100\rangle\langle100|\nonumber\\
&& +{\rm sgn}(g)(|110\rangle\langle110|+|101\rangle\langle101|+|011\rangle\langle011|), \nonumber
\end{eqnarray}
where ${\rm sgn}(g)$ is the sign function, which we take to be 1 if $g\ge 0$, and -1 if $g<0$. 
Thus, for simplicity, we only need to consider $g>0$ case.  For $0\le g\le 1$, we construct a  local POVM that contains two elements $\{ E_1=F_1^\dagger F_1, E_2=F_2^\dagger F_2\}$, where
\begin{eqnarray}
F_1&=&g(|000\rangle\langle 000|+|111\rangle\langle 111|)\nonumber\\ 
& &+|001\rangle\langle001|+|010\rangle\langle010|+|100\rangle\langle100|\nonumber\\
&& +|110\rangle\langle110|+|101\rangle\langle101|+|011\rangle\langle011|, \\
F_2&=&\sqrt{1-g^2}(|000\rangle\langle 000|+|111\rangle\langle 111|).
\end{eqnarray}
One can verify that $E_1+E_2=\openone$ and thus the POVM is trace-preserving. When the outcome is $E_1$, the local tensor is reduced back to the fixed-point form, except for the $g$ factor,
\begin{align}
& A(0,0,0) = A(1,1,1) =g, \notag\\ 
& A(0,0,1) = A(0,1,0) = A(1,0,0) = g \notag \\
& A(1,1,0) = A(1,0,1) = A(0,1,1) = g.
\end{align}
  When the outcome is $E_2$, the local tensor becomes
\begin{align}
& A(0,0,0) = A(1,1,1) = \sqrt{1-g^2}, \notag\\ 
& A(0,0,1) = A(0,1,0) = A(1,0,0) = 0 \notag \\
& A(1,1,0) = A(1,0,1) = A(0,1,1) = 0,
\end{align}
and this means that the three GHZ loops surrounding the vertex have been merged into one giant loop.

 We label the POVM outcome by $\{\alpha_{v}\}$, where $v$ labels the site and $\alpha_v$ can be either $1$ (desired) or $2$ (undesired) measurement outcome.
 We label the total number of outcome $E_1$ by $n_1$ and that of $E_2$ by $n_2$ and the size of cluster $c$ by $|c|$. The latter $|c|$ counts the number of hexagons in a loop (see Fig.~\ref{fig:loops}). We can show that the probability of a particular outcome $\{\alpha_v\}$ is proportiontal to
 \begin{equation}
 \label{eqn:appendixP}
 P_{\{\alpha_v\}}=p_0\, g^{2n_1} (1-g^2)^{n_2}\, 2^{\sum_c (1-|c|)},
 \end{equation}
 where $p_0$ is an overall constant that is independent of $\{\alpha_v\}$.
 
 To see this, we first observe that the wave function at $g=1$ corresponds to a product of (un-normalized) GHZ loops (with the number equal to the number of faces), i.e. each loop on a hexagon containing six qubits is in a state $|000000\rangle + |111111\rangle$. Away from this fixed point $g\ne 1$, the GHZ loops get distorted and become entangled through the local deformation at a vertex. The effect of $E_1$ at a vertex $v$ is to restore locally so that the three qubits are disentangled and if all six vertices around a hexagon have $E_1$ outcome then the GHZ loop in that hexagon is restored. Each $E_1$ contributes a factor of $g$ in the wave function. The effect of $E_2$ at site $v$ is to merge the three GHZ loops passing this vertex and to multiply the wave function by a factor of $\sqrt{1-g^2}$. Therefore, for any configuration of outcomes at all sites $\{\alpha_v\}$, the resulting state is a product of GHZ loops, but their number is equal to $n_F- \sum_c(|c|-1)$, where $n_F$ is the total number of faces or hexagons in the lattice. Each un-normalized GHZ loop contribute to a factor of 2 in the probability. Taking into account of $n_1$ factors of $g$ and $n_2$ factors of $\sqrt{1-g^2}$ in the amplitude, we arrive at
\begin{equation}
 P_{\{\alpha_v\}}= g^{2n_1} (1-g^2)^{n_2}\, 2^{\sum_c (1-|c|)} 2^{n_F}/\langle \psi(g)|\psi(g)\rangle,
 \end{equation}
 where $\langle \psi(g)|\psi(g)\rangle$ is the norm square of the original un-normalized wave function, which can also be omitted for the purpose of Monte Carlo sampling, and the factor involing $n_F$ can be omitted as well. Therefore, we have proved the expression in Eq.~(\ref{eqn:appendixP}).
  
  We remark that Eq.~(\ref{eqn:appendixP}) can be interpreted as the Boltzmann weight of a statistical model of interacting particles of two types:
 \begin{equation}
 P_{\{\alpha_v\}}(g)\sim e^{- n_1 \epsilon_1 - n_2 \epsilon_2 - U_2},
 \end{equation}
 where a type-1 particle costs energy $\epsilon_1\equiv \ln (1/g^2)$ and a type-2 particle costs $\epsilon_2\equiv \ln (1/(1-g^2))$, and additionally the interacton potential $U_2$ among type-2 particles due to loop stretching and merging: $-|c|\ln 2$, where $|c|$ depends on the configuration of the type-2 particles.

\subsection{Case $|g|> 1$}
\label{app:Honeycomb2}
In this region of the parameter, we consider a different POVM that contains two elements $\{ E_3=F_3^\dagger F_3, E_4=F_4^\dagger F_4\}$, where
\begin{eqnarray}
\!\!\!\!\!\!\!F_3&=&\frac{1}{g}(|001\rangle\langle001|+|010\rangle\langle010|+|100\rangle\langle100|\nonumber\\
 \!\!\!\!\!\!\!& & +|110\rangle\langle110|+|101\rangle\langle101|+|011\rangle\langle011|)\\
& &+(|000\rangle\langle 000|+|111\rangle\langle 111|)\\
\!\!\!\!\!\!\!F_4&=&{\frac{\sqrt{g^2-1}}{g}}(|001\rangle\langle001|+|010\rangle\langle010|+|100\rangle\langle100|\nonumber\\
\!\!\!\!\!\!\!& & +|110\rangle\langle110|+|101\rangle\langle101|+|011\rangle\langle011|).
\end{eqnarray}
One can verify that $E_3+E_4=\openone$ and thus the POVM is trace-preserving. When the outcome is $E_3$, the local tensor is reduced back to the fixed-point form, except for the $1/g$ factor. When the outcome is $E_4$, the local tensor becomes
\begin{align}
& A(0,0,0) = A(1,1,1) =0, \notag\\ 
& A(0,0,1) = A(0,1,0) = A(1,0,0) = \sqrt{g^2-1}\notag \\
& A(1,1,0) = A(1,0,1) = A(0,1,1) = \sqrt{g^2-1}.
\end{align}
and this means that the three GHZ loops surrounding the vertex have been merged into one giant loop that can have more than two components. For example, one single site with $E_4$ outcomes merged the three neighboring hexagon loops of GHZ state into one GHZ state of six components 
\begin{align}
& |0...0\rangle|0...0\rangle|1...1\rangle+ |1...1\rangle|1...1\rangle|0...0\rangle \nonumber\\
&+ |0...0\rangle|1...1\rangle|0...0\rangle+|1...1\rangle|0...0\rangle|1...1\rangle \nonumber\\
&+ |1...1\rangle|0...0\rangle|0...0\rangle+|0...0\rangle|1...1\rangle|1...1\rangle.
\end{align}
 To recover the case of two-component GHZ state, one can perform a measurement that consists of three projection  at the vertex (after the $E_2$ outcome): $P_0=|000\rangle\langle000|+|111\rangle\langle111|$, $P_1=|001\rangle\langle001|+|110\rangle\langle110|$, $P_2=|010\rangle\langle010|+|101\rangle\langle101|$, $P_3=|100\rangle\langle100|+|011\rangle\langle011|$. Since the measurement is done following the $E_2$ outcome, there will not be an outcome associated with $P_0$, and each outcome associated with either $P_1$, $P_2$ or $P_3$, projects to a level-two subspace that can be locally converted to $P_0$ subspace. The reduction to a universal resource state goes as before, up to local rotation. 

 We label the POVM outcome similarly by $\{\alpha_{v}\}$, where $v$ labels the site and $\alpha_v$ can be either $3$ (desired) or $4$ (undesired) measurement outcome.
 We also label the total number of outcome $E_3$ by $n_3$ and that of $E_4$ by $n_4$. We show below that the probability of a particular outcome $\{\alpha_v\}$ is proportiontal to
 \begin{equation}
 \label{eqn:appendixP'}
 P_{\{\alpha_v\}}=p_0'\, g^{-2n_3} \left(\frac{g^2-1}{g^2}\right)^{-n_4} \, \prod_{c} q_c\, 2^{-|c|} ,
 \end{equation}
 where $p_0'$ is an overall constant that is independent of $\{\alpha_v\}$, and $q_c$ counts the number of components in the merged generalized GHZ state for the cluster $c$. For the previous $g<1$ case, the $q_c=2$, but due to the more complicated structure in the $E_4$ here, $q_c$ can be larger than 2. But since there are 6 components in $E_4$, we have a lower bound $q_c\ge 6$. We also observe that  $q_c \le 6^{|c|/3}$, for which an example of saturation is $|c|=3$ for a triangle.

 The proof for Eq.~(\ref{eqn:appendixP'}) is similiar to that of Eq.~(\ref{eqn:appendixP}). Since $g>1$ we can divide the original tensor by this factor, and thus we turn the problem of $g>1$ to $1/g <1$. Thus the role of $E_1$ is now played by $E_3$ and that of $E_2$ by $E_4$, except that $E_4$ contains 6 rank-1 projectors, whereas $E_2$ contains 2 rank-1 projectors. The factor $2^{1}$ in Eq.~(\ref{eqn:appendixP}) exactly counts the two components in the GHZ loop but now the loop will contain more components of at least 6. Thus we denote the total number of components in a loop inside a cluster $c$ by $q_c$. This establishes the proof for Eq.~(\ref{eqn:appendixP'}). We do not know of a closed form for $q_c$ using simple geometric properties of the cluster, such as the number of its vertices, edges or loops. The counting has to be done case by case. Thus in our Monte Carlo simulations we will replace $q_c$ by its bounds $6\le q_c\le 6^{|c|/3}$, and at least the transition point of the quantum computational universality can be bounded.

We remark that the above analysis for the probability distribution extends to other lattices with appropriate extension of SPT wave functions, such as the square lattice or even 3D cubic or diamond lattices.

\section{Monte Carlo simulations for quantum computational universality}
\label{app:MonteCarlo}
\begin{figure}[t]
\subfigure[]{\includegraphics[width=0.49\textwidth]{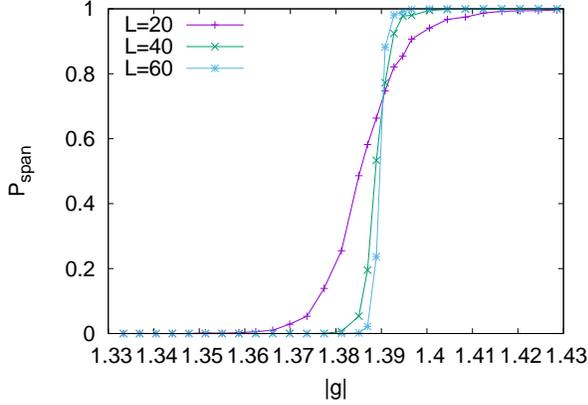}}
\subfigure[]{\includegraphics[width=0.49\textwidth]{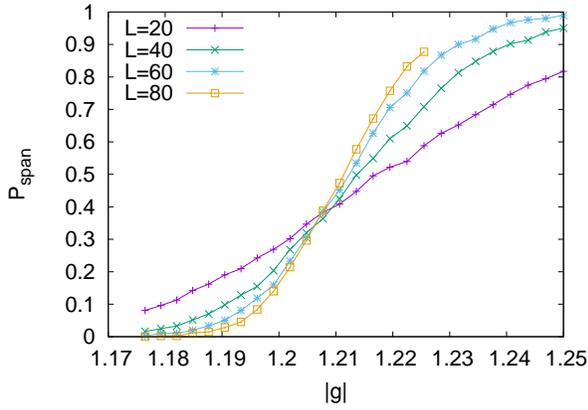}}
\caption{Probability $P_{\rm span}$ of a spanning loop (i.e. a macroscopic GHZ loop) vs. parameter $g$ for the case $g>1$ on the honeycomb lattice. The top panel gives an estimate of the upper bound whereas the bottom panels gives the lower bound on the threshold value $g_{c2}$ that separates regions with and without quantum computational universality. It is estimated that $1.205(5)\lesssim g_{c2} \lesssim 1.390(2)$.}
\label{fig:percoQ}
\end{figure}

With the derived probability for any given outcome $\{\alpha_v\}$, we can employ Monte Carlo methods. In particular, we use the Metropolis method to attempt to flip at a site (labeled by zero) from $E_1$ to $E_2$ or vice versa for $g<1$ (and between $E_3$ and $E_4$ for $g>1$), denoted by $\alpha_0\rightarrow \alpha_0'$. The probability ratio of $P_{\alpha_{v\ne 0}; \alpha_0'}/P_{\alpha_{v\ne 0};  \alpha_0}$ depends locally on $\Delta n_1$ and $\Delta n_2$, but quasi-locally on $|c|$ and $|c'|$. The cluster size can be obtained by a Wolff-like algorithm, and is the bottleneck of the simulations.  The flip is accepted with a probability
\begin{equation}
P_{\rm accept}=\min\{1, P_{\alpha_{v\ne 0}; \alpha_0'}/P_{\alpha_{v\ne 0};  \alpha_0}\}.
\end{equation}
The initialization of configuration $\{\alpha_v\}$ can be assigned randomly (i.e. hot start) or uniformly to $E_1$ or $E_2$ (cold start).

From the argument of continuity, we expect that there exist a $g_{c1}<1$ and $g_{c2}>1$ such that for $g\in (g_{c1},g_{c2})$ inside the SPT phase, the quantum computation using the SPT ground states as the resource for MBQC is universal.

We find that the location of $g_{c1}\approx 0.760(2)$ to be very close, if not identical, to the transition from the SPT phase to the $Z_2$ symmetry breaking phase. It is likely that the transition in the quantum computational power coincides with the transition to the symmetry-breaking phase. This is due to the fact that the long-range order starts to form at the beginning of a percolated phase. This suggests that the POVM is optimal for $g<1$. For the other side $g>1$, we can bound the value of $g_{c2}$: $1.205(5)\lesssim g_{c2} \lesssim 1.390(2)$; see Fig.~\ref{fig:percoQ}. However, there is no transition of phases nearby. There are two possibilities: one being that our POVM for $g>1$ is not optimal but the quantum computational power extends further beyond, or it is possible that the quantum computational power does not extend all the way to the whole SPT phase. However, we do not have the means to resolve this. Obtaining a whole SPT phase that provides universal resource for quantum computation will likely further strengthen the connection between quantum computation and the SPT order in two and higher dimensions. Such connection and whether quantum computational capability extends to the whole SPT phase may also depend on the SPT order. Characterizing such a connection would certainly reveal the fundamental connection of quantum computational power and topological symmetric phases of matter.

\section{Square lattice}
\label{app:Square}

\begin{figure}[t]
\center{\epsfig{figure=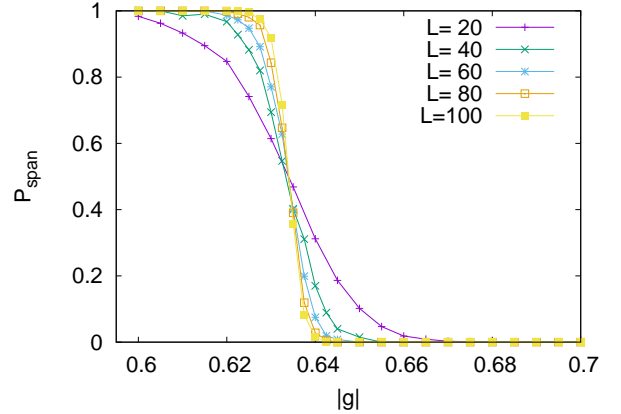,angle=0,width=.5\textwidth}}
\vspace{-0.8cm}
\caption{Probability $P_{\rm span}$ of a spanning loop (i.e. a macroscopic GHZ loop) vs. parameter $g$ for the case $g<1$ for various linear sizes $L$ on the square lattice. The total number of squares is $L^2$.  A percolation transition is estimated at $g_{c1}\approx 0.635(3)$ in the thermodynamic limit. In the region $g_{c1}< g\le 1$, there is no macroscopic GHZ loop ($P_{\rm span}\rightarrow 0$ as $L\rightarrow \infty$) and hence a sufficiently large valence-bond state with the number of qubits being proportional to the original system size can be obtained. This means that the system provides a universal resource for MBQC.}
\label{fig:percoSQ}
\end{figure} 
Here we present the construction of the POVM, the outcome probability and the simulation results for the square lattice.

\subsection{POVM outcome probability for $|g|\le 1$}

\label{app:Square1}
Since the phase is local to each site of three qubits, for the purpose of quantum computation we can transform it away by local unitary. Thus, for simplicity, we only need to consider $g>0$ case. In this part we first discuss the parameter range $0\le g\le 1$. In this region, we construct a  local POVM that contains two elements $\{ E_1=F_1^\dagger F_1, E_2=F_2^\dagger F_2\}$, where
\begin{eqnarray}
F_1&=&g(|0000\rangle\langle 0000|+|1111\rangle\langle 1111|)+\sum_{\rm rest} |{\rm rest}\rangle\langle{\rm rest}| \nonumber \\
F_2&=&\sqrt{1-g^2}(|0000\rangle\langle 0000|+|1111\rangle\langle 1111|),
\end{eqnarray}
where $|{\rm rest}\rangle\langle {\rm rest}|$ denotes product projectors other than those associated with $0000$ and $1111$.
One can verify that $E_1+E_2=\openone$ and thus the POVM is trace-preserving. 
The argument used in the honeycomb case  can be repeated to show that the probability of a particular outcome $\{\alpha_v\}$ is proportiontal to
 \begin{equation}
 \label{eqn:appendixSqP}
 P_{\{\alpha_v\}}=p_0\, g^{2n_1} (1-g^2)^{n_2}\, 2^{\sum_c (1-|c|)},
 \end{equation}
 where $p_0$ is an overall constant that is independent of $\{\alpha_v\}$.
 
Using this we have performed Monte Carlo simulations and obtain the quantum computational universality disappears at  $g_c\approx 0.635(3)$ when $g$ decreases from 1; Fig.~\ref{fig:percoSQ}.
 \subsection{Case $|g|> 1$}
 \label{app:Square1}
 In this region of the parameter, we consider a different POVM that contains two elements $\{ E_3=F_3^\dagger F_3, E_4=F_4^\dagger F_4\}$, where
\begin{eqnarray}
F_3&=&\frac{1}{g}\sum_{\rm rest} |{\rm rest}\rangle\langle{\rm rest}|+ |0000\rangle\langle 0000|+|1111\rangle\langle 1111|\nonumber \\
 F_4&=&{\frac{\sqrt{g^2-1}}{g}}\sum_{\rm rest} |{\rm rest}\rangle\langle{\rm rest}|.
\end{eqnarray}
One can verify that $E_3+E_4=\openone$ and thus the POVM is trace-preserving.
Similar to the honeycomb case, the probability of a particular outcome $\{\alpha_v\}$ can be shown to be 
 \begin{equation}
 \label{eqn:appendixSqP'}
 P_{\{\alpha_v\}}=p_0'\, g^{-2n_3} \left(\frac{g^2-1}{g^2}\right)^{-n_4} \, \prod_{c} q_c\, 2^{-|c|} ,
 \end{equation}
 where $p_0'$ is an overall constant that is independent of $\{\alpha_v\}$, and $q_c$ counts the number of components in the merged generalized GHZ state for the cluster $c$. Since the there are four partons on a site, the $q_c$ has slightly different bounds:  $14\le q_c \le 14^{|c|/4}$.
 
 Using this we have performed Monte Carlo simulations and obtain the quantum computational universality disappears at  $g_{c2}$, where
 $1.31(1)\lesssim g_{c2}\lesssim 1.82(1)$.

\section{$Z_2$ SPT order and modular S and T matrices}
\label{app:ST}
Here we show the explict form of the modular S and T matrices from our numerical results.
\subsection{Honeycomb case}
When $-0.760(2) \leq g < 0.760(2)$, the modular matrices are
\begin{align}
\label{T2matrix_tri}
S = T= \begin{pmatrix}
  1 & 0&0 &0 \\
  0& 0&0 &0 \\
   0 & 0&0 &0\\
   0 & 0&0 &0
 \end{pmatrix},
\end{align}
and they show that it is a symmetry-breaking phase.  
When $g>0.760(2)$,  the modular matrices are
\begin{align}
\label{tri_Z2}
S =  \begin{pmatrix}
  1 & 0&0 &0 \\
  0& 0&1 &0 \\
   0 & 1&0 &0\\
   0 & 0&0 &1
 \end{pmatrix},  \ \
 T =  \begin{pmatrix}
  1 & 0&0 &0 \\
  0& 1&0 &0 \\
   0 & 0&0 &1\\
   0 & 0&1 &0
 \end{pmatrix}.
\end{align}
One can diagonalize the $T$ matrix and obtain the corresponding $S$ matrix (also enforcing the elements in first row and the first column to be $1/2$)
\begin{align}
\label{tri_Z2prime}
T' =  \begin{pmatrix}
  1 & 0&0 &0 \\
  0& 1&0 &0 \\
   0 & 0&1 &0\\
   0 & 0&0 &-1
 \end{pmatrix}, \\
S' =\frac{1}{2}  \begin{pmatrix}
  1 & 1&1 &1 \\
  1& 1&-1 &-1 \\
   1 & -1&1 &-1\\
   1 & -1&-1 &1
 \end{pmatrix},  
\end{align}
which are the modular matrices of the toric code.
In this region, ${\rm Tr}(T^2)=4$, showing    the trivial $\mathbb{Z}_2$ SPT order ($\text{SPT}^0$).
For  $g<-0.760(2)$, the modular matrices are
\begin{align}
\label{nontri_Z2}
S =  \begin{pmatrix}
  1 & 0&0 &0 \\
  0& 0&1 &0 \\
   0 & 1&0 &0\\
   0 & 0&0 &-1
 \end{pmatrix},  \ \
 T =  \begin{pmatrix}
  1 & 0&0 &0 \\
  0& 1&0 &0 \\
   0 & 0&0 &-1\\
   0 & 0&1 &0
 \end{pmatrix},
\end{align}
with ${\rm Tr}(T^2)=0$. This shows a nontrivial $\mathbb{Z}_2$ SPT order  ($\text{SPT}^1$). Similarly, we can diagonalize $T$ and transform $S$ to the same basis:
\begin{align}
\label{nontri_Z2prime}
T' =  \begin{pmatrix}
  1 & 0&0 &0 \\
  0& 1&0 &0 \\
   0 & 0&i &0\\
   0 & 0&0 &-i
 \end{pmatrix}, \ \
S' =  \frac{1}{2}\begin{pmatrix}
  1 & 1&1 &1 \\
  1& 1&-1 &-1 \\
   1 & -1&-1 &1\\
   1 & -1&1 &-1
 \end{pmatrix},
\end{align}
which are the modular matrices for the double-semion model~\cite{LevinGu2012}.
\subsection{Square case}
The modular matrices have the same forms in the respective symmetry-breaking, trivial SPT and nontrivial SPT phases. The only difference is the range of the parameter $g$.
\end{document}